\documentclass{pazhastl}
\usepackage{graphicx}
\usepackage{url}
\begin{document}

\journalinfo{2010}{36}{02}{001}[000]

\title{Automated search for star clusters in large multiband surveys: II. Discovery and investigation of open clusters in the Galactic plane}

\author{E.~V.~Glushkova\address{1}\email{elena@sai.msu.ru}, S.~E.~Koposov\address{1,2,3}, I.~Yu.~Zolotukhin\address{1}, Yu.~V.~Beletsky\address{4}, A.~D.~Vlasov\address{1}, S.~I.~Leonova\address{1}
\small
\\
\vskip 3mm
$^1$ {\it Sternberg Astronomical Institute, Moscow, Russia}\\
$^2$ {\it Max Planck Institute for Astronomy, Heidelberg, Germany}\\
$^3$ {\it Institute of Astronomy, University of Cambridge, UK}\\
$^4$ {\it European Southern Observatory, Santiago, Chile}\\
  }

\shortauthor{GLUSHKOVA ET AL.}

\shorttitle{AUTOMATED SEARCH FOR STAR CLUSTERS}

\submitted{September 20, 2009; in final form, September 25, 2009}

\begin{abstract}
Automated search for star clusters in $J,H,K_s$ data from 2MASS catalog has been performed using the method developed by Koposov et. al (2008). We have found and verified 153 new clusters in the interval of the galactic latitude $-24^{\circ}<b<24^{\circ}$. Color excesses $E(B-V)$, distance moduli and ages were determined for 130 new and 14 yet-unstudied known clusters. In this paper, we publish a catalog of coordinates, diameters, and main parameters of all the clusters under study. A special web-site available at \url{http://ocl.sai.msu.ru} has been developed to facilitate dissemination and scientific usage of the results.

\end{abstract}

\section{INTRODUCTION}

The search for star clusters in the Galaxy is of a great interest for investigators in the recent years. On the one hand, the usage of new methods and instruments in observations allows scientists to involve clusters in solving a larger number of astrophysical problems and raises many new ones, such as: the processes of formation and evolution of young massive clusters, the nature of nuclear star clusters, the presence of multiple populations in globular clusters. On the other hand, the availability of new multiband all-sky surveys stimulates us to search them for new clusters and build a homogeneous catalog of parameters of both newly discovered and already known clusters. 

In Paper I (Koposov et al. 2008), we described the new method of an automated search for star clusters as a density peaks in huge stellar catalogs. It is based on the convolution of density maps with a special 2-D filter, which is the difference between two 2-D Gaussian profiles and has zero integral. If convolved with this special filter, the areas of flat or slowly changing background would produce zero signal, whereas the areas of star concentrations would exhibit a high signal. Using this method, we analyzed the distribution of stars in Two Micron All Sky Survey (2MASS) in the field of $16\times16$ degrees towards the Galactic anticenter and found 15 new open clusters. To verify the reality of detected clusters, we developed a method, which is based on the assumption that probable cluster members lying along the same isochrone on the color-magnitude diagram (CMD) should also form the spatial density peak, whereas the background stars should have a flat distribution. The other benefit of this method is that it not only allows to verify a cluster candidate, but at the same time estimates the cluster's main parameters -- distance, age, and color excess. We employed this method to derive parameters of 12 new and 13 known, but poorly studied clusters, which were also detected in the field of interest, using the technique of $(J,J-H)$ and $(K_s,J-K_s)$ diagrams built with the data from 2MASS catalog. Later, $UBVI$  magnitudes of the stars in three new clusters were measured using CCD images taken at 104-cm telescope of Aryabhatta Research Institute of Observational Sciences (ARIES, India). The isochrones fitted to these data using another color-magnitude diagrams, and derived clusters parameters (Glushkova et al. 2009) show good agreement with those obtained from $(J,J-H)$ and $(K_s,J-K_s)$ 2MASS diagrams and therefore independently confirm the reality of clusters found by Koposov et al. (2008) and the correctness of the method of new cluster verification.

The aims of this study are: (1) search for overdensities in the Galactic plane in the range of the galactic latitude $|b|<24^{\circ}$ using 2MASS data; (2) verification of some of them as a real clusters; (3) estimate of their physical parameters using proven technique we have developed earlier.

\section{SEARCH FOR NEW CLUSTERS AND DETERMINATION OF THEIR PARAMETERS}

\begin{figure*}
\resizebox{\hsize}{!}{\includegraphics{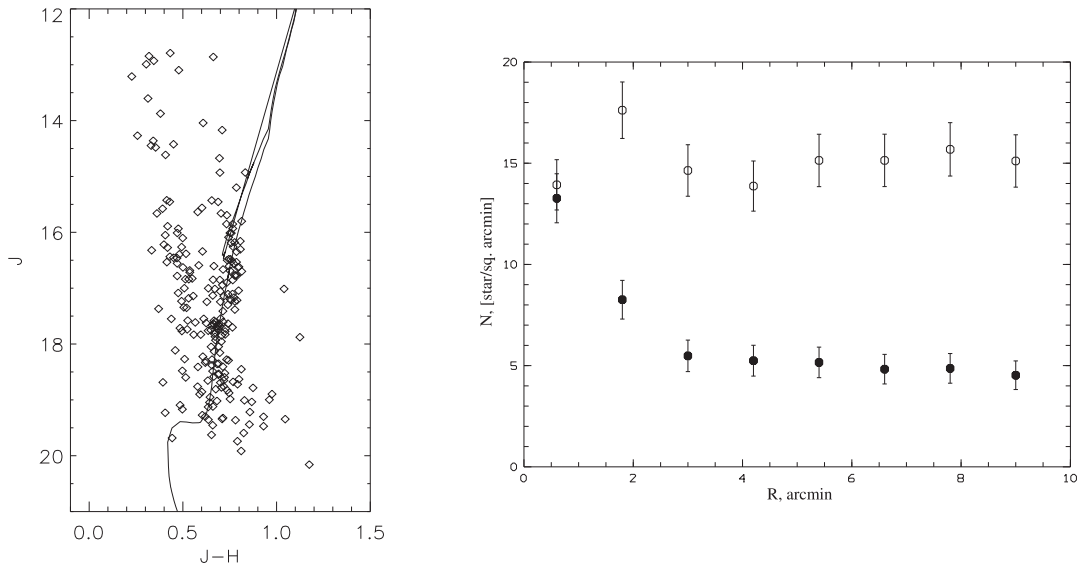}} \caption{Left panel: $(J,J-H)$ diagram of the stars within $3^{\prime}$ from the center of the SAI 50 built by data from UKIDSS GPS with the fitted isochrone. 
Right panel: Radial density distribution. Solid circles represent the density of cluster members lying along the fitted isochrone not far than $0^m.05$ in $(J-H)$ direction. Open circles denote the density of the field stars.
}
\end{figure*}

For the new clusters search and verification procedures we made use of \url{http://vo.astronet.ru}, the Virtual Observatory (VO) resource of the Sternberg Astronomical Institute (Koposov et al. 2007), which provides fast remote access to 2MASS catalog data through the standard VO-compliant interfaces. We detected 11186 overdensities with the significance level of more than $4.5\times \sigma$ in the region of the Milky Way within the interval of the galactic latitude $-24^{\circ}<b<24^{\circ}$. Since the open clusters subsystem concentrates to the Galactic plane where the interstellar extinction reaches maximum and the background changes rapidly, the most from 11186 overdensities should be attributed to the fluctuations of interstellar dust especially towards the Galactic center, where the density fluctuations are maximal. We undertook a visual inspection of all detected overdensities by images from Digitized Sky Survey (DSS) and 2MASS catalogs to recognize real density concentrations against these background fluctuations. We selected 962 candidates to clusters and matched them to the online catalogs by Dias et al. (2002) and Harris (2003). This match demonstrated that 565 candidates are known open clusters, 114 are known globular clusters, whereas 283 overdensities should be examined in detail as a new cluster candidates, some of which may only be attributed to random star concentrations. We ran all these 283 overdensities through our verification procedure, which includes the analysis of their Hess-diagrams, color-magnitude diagrams and radial density distributions in accordance with the technique by Koposov et al. (2008). 149 candidates were decided to be the real star clusters.

However, a part of non-selected overdensities may also be real clusters, which we were unable to verify, because they were detected at the sensitivity limit of 2MASS catalog. If we extend our data pattern with the data from the sources with a
higher limit in $J,H,K$ magnitudes, then some other overdensities may manifest themselves as real clusters, and their parameters can also be determined. We tested this idea by using Galactic Plane Survey of UKIRT Infrared Deep Sky Survey, Data Release 3 (UKIDSS GPS DR3, Warren et al. 2007) as an additional data source. The average 2MASS sensitivity in $K_s$ is $14^m.3$, whereas in UKIDSS GPS, the magnitude limit for $K$-band is as high as $19^m$. 22 cluster candidates found in 2MASS were studied by $(J,J-H)$ and $(K,J-K)$ diagrams and radial density distributions built using data from UKIDSS GPS. We verified 9 star clusters and determined their distance moduli, ages and color excesses. Among these nine objects, four (SAI 50, SAI 131, SAI 133, and SAI 141 in Table 1) did not reveal themselves as star clusters when only 2MASS catalog data were employed, because the upper part of their main sequences and/or red giant branches is only seen on CMD and Hess-diagrams. For example, left panel of Fig.~1 displays $(J,J-H)$ diagram of cluster SAI 50 built by data from UKIDSS GPS. One can only see here the red-giant branch limited to the stars fainter than $J=16^m$. If we build such a CMD for this cluster with 2MASS data only, we would not detect the cluster at all, because the limiting magnitude for $J$-band in this catalog is exactly $16^m$. In Fig.~1, the isochrone is fitted to give the lower estimate of the distance to the cluster.

Thus, after involving UKIDSS GPS data, the number of confirmed clusters reached 153. For 130 clusters we automatically obtained main physical parameters: ages, distances and color excesses using the data from 2MASS or UKIDSS GPS and isochrones of solar metallicity by Girardi et al. (2002). $J,H,K$ magnitudes were taken from the deeper UKIDSS GPS for the nine clusters, whose centers fall into this survey: SAI 50, SAI 74, SAI 75, SAI 130, SAI 131, SAI 133, SAI 141, SAI 142, SAI 145. The center coordinates, diameters, color excesses $E(B-V)$, distance moduli and ages as $log(t)$ are listed in Table 1. We consider the position of the maximum in a density peak as the center of a cluster on the density map. The cluster radius is such a distance from its center, at which the star density becomes flat on the radial density distribution plot. The errors in color excesses, distance moduli and ages were estimated as described by Koposov et al. (2008): from the differences in the parameters derived by $(J,J-H)$ and $(K_s,J-K_s)$ diagrams. We used the relations $A_{K_s}=0.670\times E(J-K_s)$, $A_J=0.276\times A_V$, $E(J-H)=0.33\times E(B-V)$ from the paper by Dutra et al. (2002) to derive the distance modulus and color excess $E(B-V)$, and the relation $A_K=0.626\times E(J-K)$ from the paper by He et al. (1995) for $K$ magnitudes from UKIDSS GPS. For some clusters, we are able to estimate the upper limit of the age only, in particular, when their color-magnitude diagrams do not contain red-giant stars.
Besides the cluster name SAI (\textit{S}ternberg \textit{A}stronomical \textit{I}nstitute), other names from papers by Kronberger et al. (2006) and by Froebrich et al. (2007) are used in Table 1. As mentioned by Koposov et al. (2008), the authors of both papers published the list of probable cluster candidates where further investigation is necessary to clarify their nature. Some of the overdensities independently found by us coincide with the cluster candidates from papers by Kronberger et al. (2006) and Froebrich et al. (2007). We performed a thorough analysis of these matching candidates and confirmed some of them as the real clusters. However, remaining candidates turned to be mere random star concentrations or background fluctuations (we do not publish this list here). All clusters from Table 1 having the other name according to Kronberger et al. (2006) or Froebrich et al. (2007), are listed neither in the database of open clusters (WEBDA) by Paunzen, Mermilliod (2009), nor in the catalog by Dias et al. (2002). That is why we consider them as a new clusters.

Special attention should be given to the cluster SAI 92 from Table 1. This cluster is situated $7^{\prime}$ to the south-east of NGC 2645, and both clusters feature approximately the same values of parameters. According to WEBDA, the distance from Sun to NGC 2645 equals to 1668 pc; we estimate the distance to the cluster SAI 92 to be $1580\pm130$ pc. Color excess $E(B-V)$ in the direction to NGC 2645 is $0^m.380$, and to SAI 92, $0^m.39\pm0^m.10$. Both clusters are pretty young, but differ in $log(t)$ by 0.72: WEBDA estimates the age of NGC 2645 as 7.283, whereas the present study gives $8.00\pm0.05$ for SAI 92. However, 0.05 is the formal error, which, as mentioned earlier, should be attributed to the difference in the ages determined by $(J,J-H)$ and $(K_s,J-K_s)$ diagrams. A real error for age estimate by 2MASS is not less than 0.3 for clusters which do not contain stars on the red-giant branch. If we build the $(V,B-V)$ diagram of NGC 2645 using data from WEBDA and fit them by the isochrone of solar metallicity by Girardi et al. (2002) using the distance and color excess also from WEBDA, then the lower limit for the age is 7.6 in $log(t)$. Therefore, the age of both clusters can be considered approximately the same taking into account the real errors of its estimate. Apparently, NGC 2645 and SAI 92 form a double cluster then. Fig.~2 shows the image of $15^{\prime}$ x $15^{\prime}$ from DSS where both clusters fall into. Diameters of clusters ($5^{\prime}$ and $3^{\prime}$) are represented by circles.

SAI 50 looks like a globular cluster by its age and CMD built using data from UKIDSS GPS, although the errors in parameters are large, because the isochrone was fitted to the red-giant branch only. The left panel of Fig.~1 represents $(J,J-H)$ diagram with fitted isochrone; the radial density distribution corresponding to this fitted isochrone is displayed in the right panel. Solid circles denote the cluster members that are the stars deviating from the isochrone by less than $0^m.05$ in color; open circles denote background stars, that are all other stars. It can clearly be seen that the members of the cluster are concentrated to its center, whereas field stars exhibit flat distribution with small density fluctuations. 

Besides 130 clusters which are listed in Table 1, we found 23 new embedded clusters. They are detected as density peaks and are clearly visible on 2MASS images and Hess-diagrams. But such a clusters reveal themselves as a cloud on the CMD $(J,J-H)$ and $(K_s,J-K_s)$; therefore, it was impossible to fit isochrones and find their parameters. Coordinates and diameters of these clusters are presented in Table 2.

\begin{table*}[t]

\vspace{0mm} \centering {{\bf Table 1.} Parameters of new clusters}\label{meansp}
\footnotesize

\vspace{5mm}\begin{tabular}{l|l|c|c|c|c|c|c} \hline\hline
{Name} & 
{Other name} &
$\alpha_{J2000}$ &
 $\delta_{J2000}$& 
{d} & 
{E(B-V)} &  
{$(m-M)_0$} &  
{Age}  
\\
&
 &
{$h$  $m$  $s$} & 
{$\circ$  $\prime$  $\prime \prime$} & 
{$\prime$} & 
{mag} & 
{mag} & 
{$log(t)$}

\\
\hline

SAI 1&              & 00:08:20.4& +51:43:15 & 4& 0.34$\pm$0.14& 11.68$\pm$0.06& 9.10$\pm$0.05 \\ 
SAI 2&              & 00:12:28.4& +76:14:06 & 5& 0.81$\pm$0.27& 12.45$\pm$0.26&  $<$7.70 \\
SAI 3&   FSR 480    & 00:14:49.6& +61:28:36 & 6& 0.68$\pm$0.02& 14.18$\pm$0.05& 9.20$\pm$ 0.05 \\
SAI 4&              & 00:23:40.0& +62:42:14 & 4& 0.27$\pm$ 0.18&  11.02$\pm$ 0.36& $<$8.70 \\
SAI 5&   FSR 494    & 00:25:37.8& +63:45:40 & 4& 0.60$\pm$ 0.08&  12.70$\pm$ 0.07& 9.10$\pm$ 0.05\\
SAI 6&              & 00:27:52.6& +60:41:21 & 5& 0.89$\pm$ 0.09&  12.17$\pm$ 0.01& $<$8.50\\
SAI 7&   FSR 503    & 00:29:11.3& +68:56:11 & 7& 0.82$\pm$ 0.12&  11.88$\pm$ 0.38& 9.40$\pm$ 0.05\\
SAI 8&   FSR 505    & 00:40:08.8& +61:34:03 & 3& 0.30$\pm$ 0.11&  14.28$\pm$ 0.20& 9.15$\pm$ 0.10\\
SAI 9&   FSR 508    & 00:41:51.9& +64:58:51 & 4& 0.66$\pm$ 0.17&  14.42$\pm$ 0.20& 9.10$\pm$ 0.10\\
SAI 11&  FSR 524    & 00:57:12.4& +62:06:18 & 8& 0.51$\pm$ 0.13&  10.70$\pm$ 0.29& 9.10$\pm$ 0.05\\
SAI 12&             & 01:04:24.1& +45:36:25 & 4& 1.13$\pm$ 0.30&  12.06$\pm$ 0.02& $<$8.15\\
SAI 13&  FSR 536    & 01:19:37.6& +63:03:20 & 8& 1.27$\pm$ 0.23&  12.38$\pm$ 0.17& $<$8.20\\
SAI 14&             & 01:26:00.3& +62:37:33 & 4& 0.55$\pm$ 0.03&  14.05$\pm$ 0.03& 9.10$\pm$ 0.05 \\
SAI 16&             & 02:05:29.6& +62:15:54 & 5& 0.70$\pm$ 0.01&  14.42$\pm$ 0.20& 9.15$\pm$ 0.05 \\
SAI 17&             & 02:20:48.3& +59:11:06 & 2& 0.98$\pm$ 0.17&  12.88$\pm$ 0.10& 8.90 $\pm$ 0.05 \\
SAI 21&             & 02:34:45.4& +62:34:22 & 3& 0.58$\pm$ 0.01&  11.35$\pm$ 0.10& $<$8.50\\
SAI 25&             & 03:00:26.7& +57:16:02 & 7& 0.84$\pm$ 0.13&  11.80$\pm$ 0.09& 9.15$\pm$ 0.05 \\
SAI 26&             & 03:09:30.8& +56:47:26 & 3& 1.47$\pm$ 0.29&  11.79$\pm$ 0.51& $<$8.20 \\
SAI 27&             & 03:11:21.9& +68:54:01 & 5& 0.62$\pm$ 0.04&  10.48$\pm$ 0.28& 9.25$\pm$ 0.10\\
SAI 29&  FSR 634    & 03:40:21.5& +59:24:52 & 5& 0.84$\pm$ 0.01&  12.24$\pm$ 0.07& 9.00$\pm$ 0.05\\
SAI 30&             & 03:40:22.2& +82:13:60 & 3& 1.25$\pm$ 0.44&  12.46$\pm$ 0.05& $<$8.00 \\
SAI 31&             & 03:51:15.4& +58:46:08 & 6& 0.16$\pm$ 0.22&  12.12$\pm$  0.02& 9.60$\pm$  0.05\\
SAI 33& Juchert 19  & 04:02:20.2& +52:26:39 & 2& 0.28$\pm$ 0.01&  10.64$\pm$ 0.02& $<$9.30 \\
SAI 34&  FSR 660    & 04:02:37.8& +51:55:50 & 5& 1.20$\pm$ 0.12&  13.78$\pm$  0.03& 9.10$\pm$  0.05\\
SAI 35& Juchert 20  & 04:10:47.0& +46:52:01 & 5& 0.70$\pm$ 0.18&  11.92$\pm$ 0.02& $<$8.55 \\
SAI 36&             & 04:16:50.4& +41:04:00 & 3& 0.55$\pm$ 0.05&  12.36$\pm$  0.19& 9.10$\pm$  0.05\\
SAI 37&  FSR 687    & 04:39:27.1& +48:12:23 & 4& 0.77$\pm$ 0.03&  14.30$\pm$  0.04& 9.15$\pm$  0.05\\
SAI 40&             & 05:00:56.9& +41:14:02 & 8& 0.35$\pm$ 0.12&  13.29$\pm$  0.05& 8.75$\pm$  0.05\\
SAI 41&             & 05:07:05.4& +38:25:05 & 2& 0.51$\pm$ 0.18&  13.58$\pm$  0.23& 9.30$\pm$  0.05\\
SAI 42& Juchert 23  & 05:07:39.7& +17:36:00 & 8& 0.62$\pm$ 0.18&  11.67$\pm$  0.09& 8.70$\pm$  0.05\\
SAI 43&             & 05:08:16.6& +49:52:08 & 2& 0.18$\pm$ 0.03&  12.92$\pm$  0.07& 8.95$\pm$  0.05\\
SAI 44&  FSR 716    & 05:11:07.4& +45:43:09 & 5& 0.24$\pm$ 0.01&  12.64$\pm$  0.06& 8.95$\pm$  0.05\\
SAI 45&  FSR 727    & 05:16:35.0& +45:34:56 & 6& 0.25$\pm$ 0.02&  10.55$\pm$  0.17& 9.20$\pm$  0.05\\
SAI 46&             & 05:19:37.0& +36:30:32 & 3& 0.78$\pm$ 0.01&  13.16$\pm$  0.12& 9.45$\pm$  0.05\\
SAI 47&             & 05:23:58.0& +42:18:52 & 3& 0.60$\pm$ 0.01&  12.93$\pm$  0.05& 8.50$\pm$  0.05\\
SAI 48&             & 05:24:16.8& +33:30:02 & 4& 0.31$\pm$ 0.07&  11.30$\pm$  0.20& 9.35$\pm$  0.05\\
SAI 49&             & 05:26:24.9& +50:47:03 & 6& 0.47$\pm$ 0.01&  12.17$\pm$  0.17& 9.05$\pm$  0.05\\
SAI 50&             & 05:28:38.9& +35:01:19 & 3& 0.60$\pm$ 0.30&  16.70$\pm$  0.50& 9.70$\pm$  0.20\\
SAI 51&             & 05:30:01.9& +33:25:31 & 3& 0.62$\pm$  0.18&  9.54$\pm$  0.20& $<$ 9.30\\
SAI 56&  FSR 852    & 05:53:30.6& +25:10:41 & 8& 0.43$\pm$  0.07&  11.82$\pm$  0.04& 8.90$\pm$  0.05\\
SAI 57&  FSR 932    & 06:04:21.5& +14:34:22 & 5& 0.39$\pm$  0.07&  10.18$\pm$  0.10& 8.90$\pm$  0.10\\
SAI 58&  FSR 921    & 06:05:09.4& +16:41:03 & 5& 1.33$\pm$  0.33&  11.80$\pm$  0.20& 8.50$\pm$  0.05\\
SAI 59&  FSR 942    & 06:05:59.2& +13:39:52 & 8& 0.59$\pm$  0.01&  11.51$\pm$  0.25& 9.00$\pm$  0.05\\
SAI 60&             & 06:06:38.6& +15:56:51 & 4& 0.90$\pm$  0.01&  14.50$\pm$  0.05& 9.05$\pm$  0.05\\
SAI 61&  FSR 904    & 06:06:55.2& +19:00:26 & 8& 0.14$\pm$  0.16&  10.51$\pm$  0.02& 8.80$\pm$  0.05\\
SAI 62&  FSR 985    & 06:11:49.0& +07:01:22 & 3& 0.55$\pm$  0.02&  12.10$\pm$  0.09& $<$ 8.60 \\
SAI 63&  FSR 987    & 06:13:44.5& +06:56:58 & 4& 0.37$\pm$  0.13&  11.40$\pm$  0.09& 8.65$\pm$  0.05\\
SAI 64&  FSR 948    & 06:25:55.9& +15:51:39 & 4& 0.08$\pm$  0.13&  11.84$\pm$  0.17& 9.00$\pm$  0.05\\
SAI 65&  FSR 979    & 06:31:16.2& +11:04:38 & 3& 1.17$\pm$  0.17&  11.65$\pm$  0.05& 8.45$\pm$  0.05\\
SAI 66&  FSR 1063   & 06:34:37.9& -00:16:01 & 5& 0.58$\pm$  0.01&  11.47$\pm$  0.14&$<$ 8.60\\
SAI 67&  Teutsch59  & 06:43:49.0& -00:52:60 & 6& 0.57$\pm$  0.14&  12.34$\pm$  0.04& $<$ 8.90\\
SAI 68& Patchick 90 & 06:44:42.7& -00:32:22 & 3& 0.41$\pm$  0.09&  14.00$\pm$  0.20& 8.95$\pm$  0.05\\
SAI 69&             & 06:45:27.8& -11:47:40 & 3& 0.35$\pm$  0.07&  14.27$\pm$  0.10& 9.10$\pm$  0.05\\
SAI 70&  FSR 1125   & 06:50:26.6& -05:26:08 & 4& 0.39$\pm$  0.01&  13.99$\pm$  0.09& $<$9.10\\
SAI 71&  FSR 1085   & 06:53:11.2& -00:11:31 & 4& 0.08$\pm$  0.09&  10.62$\pm$  0.19& $<$8.60\\
SAI 72&             & 06:55:48.4& +00:13:37 & 5& 0.82$\pm$  0.06&  12.49$\pm$  0.05& 8.50$\pm$  0.20\\
SAI 73&  FSR 1171   & 07:00:13.1& -10:19:14 & 2& 0.65$\pm$  0.07&  13.07$\pm$  
0.29& 9.50$\pm$  0.05\\
SAI 74&  FSR 1150   & 07:04:36.6& -06:36:38 & 3& 0.94$\pm$  0.06&  13.60$\pm$  0.26& $<$8.30\\
SAI 75& Patchick 79 & 07:15:22.5& -07:25:20 & 4& 0.23$\pm$  0.03&  12.24$\pm$  0.24& $<$8.60\\
SAI 76&  FSR 1253   & 07:26:17.4& -18:25:55 & 6& 0.62$\pm$  0.01&  12.11$\pm$  0.24& 8.40$\pm$  0.05\\
SAI 77&  FSR 1275   & 07:28:06.2& -21:47:46 & 4& 0.60$\pm$  0.10&  12.87$\pm$  0.12& 9.40$\pm$  0.05\\
SAI 78&  Teutsch 61 & 07:34:39.8& -19:47:23 & 3& 0.55$\pm$  0.02&  11.42$\pm$  0.10& 6.95$\pm$ 0.05 \\
SAI 79&  FSR 1347   & 07:48:03.4& -33:42:35 & 6& 1.09$\pm$  0.15&  11.95$\pm$  0.07& $<$8.80 \\
SAI 80&  Teutsch 25 & 07:48:26.7& -27:54:51 & 5& 0.57$\pm$  0.12&  11.48$\pm$  0.16& $<$9.00\\

\hline \multicolumn{8}{l}{}\\
\end{tabular}
\end{table*}

\begin{table*}[t]

\vspace{0mm} \centering {{\bf Table 1 (continued).} Parameters of new clusters }\label{meansp}
\footnotesize

\vspace{5mm}\begin{tabular}{l|l|c|c|c|c|c|c} \hline\hline
{Cluster} & 
{Other name} &
$\alpha_{J2000}$ &
$\delta_{J2000}$& 
{d} & 
{E(B-V)} &  
{$(m-M)_0$} &  
{Age}  
\\
&
 &
{$h$  $m$  $s$} & 
{$\circ$  $\prime$  $\prime \prime$} & 
{$\prime$} & 
mag&
mag&
{$log(t)$}

\\
\hline

SAI 81&             & 07:52:07.2& -28:07:21 & 7& 0.40$\pm$  0.02&  12.21$\pm$ 0.19& $<$ 9.20\\
SAI 82&             & 07:52:14.4& -33:02:27 & 3& 0.39$\pm$  0.07&  10.30$\pm$  0.05& 8.65 $\pm$ 0.05 \\
SAI 83&             & 07:53:20.1& -32:33:30 & 3& 1.00$\pm$  0.15&  12.96$\pm$  0.34& 8.85$\pm$  0.05\\
SAI 84&Kronberger 85& 07:58:21.6& -34:46:09 & 2& 1.17$\pm$  0.19&  14.65$\pm$  0.11& 8.80$\pm$  0.05\\
SAI 85&  FSR 1378   & 08:01:11.5& -40:40:41 & 6& 1.21$\pm$  0.25&  11.74$\pm$  0.13& 8.80$\pm$  0.05\\
SAI 86&             & 08:08:15.0& -36:36:33 & 6& 0.69$\pm$  0.09&  12.41$\pm$  0.16& 8.60$\pm$  0.05\\
SAI 87&  FSR 1380   & 08:11:05.0& -39:59:51 & 4& 1.28$\pm$  0.30&  11.80$\pm$  0.19& $<$8.65\\
SAI 88&             & 08:12:57.6& -41:53:01 & 2& 0.92$\pm$  0.02&  14.58$\pm$  0.25& 9.00$\pm$  0.05\\
SAI 89&  FSR 1387   & 08:24:25.3& -39:56:11 & 3& 1.19$\pm$  0.11&  11.82$\pm$  0.18& $<$8.80\\
SAI 90&             & 08:27:58.9& -41:46:10 & 7& 1.02$\pm$  0.21&  12.19$\pm$  0.11& 9.05$\pm$  0.05\\
SAI 91&             & 08:37:03.4& -50:03:52 & 4& 0.51$\pm$  0.02&  12.45$\pm$  0.16& 8.80$\pm$  0.05\\
SAI 92&  FSR 1436   & 08:39:34.5& -46:17:54 & 5& 0.39$\pm$  0.10&  10.99$\pm$  0.17& 8.00$\pm$  0.05\\
SAI 93&  FSR 1415   & 08:40:21.9& -44:44:01 & 3& 1.87$\pm$  0.35&  12.74$\pm$  0.06& 9.25$\pm$  0.05\\
SAI 94&             & 08:44:39.8& -46:17:46 & 5& 0.48$\pm$  0.05&  12.96$\pm$  0.29& 9.10$\pm$  0.10\\
SAI 95&  FSR 1454   & 08:51:00.6& -48:27:37 & 5& 0.57$\pm$  0.03&  13.96$\pm$  0.05& 9.15$\pm$  0.05\\
SAI 96&  FSR 1430   & 08:51:49.4& -44:15:47 & 4& 2.50$\pm$  0.33&  12.27$\pm$  0.33& 8.85$\pm$  0.05\\
SAI 98&  FSR 1435   & 08:57:03.0& -43:45:40 & 6& 1.06$\pm$  0.06&  10.27$\pm$  0.30& 9.75$\pm$  0.10\\
SAI 99&             & 08:58:04.4& -43:24:38 & 5& 1.84$\pm$  0.49&  12.60$\pm$  0.04& 8.95$\pm$  0.05\\
SAI 100& FSR 1450   & 08:58:12.2& -46:17:19 & 3& 1.62$\pm$  0.08&  13.65$\pm$  0.08& 9.10$\pm$  0.15\\
SAI 101&            & 08:58:22.4& -43:07:35 & 5& 1.56$\pm$  0.04&  11.09$\pm$  0.05& $<$9.00\\
SAI 102& FSR 1402   & 09:05:43.2& -37:52:05 & 4& 0.08$\pm$  0.04&  12.03$\pm$  0.29& 9.60$\pm$  0.05\\
SAI 103& FSR 1460   & 09:07:38.5& -47:50:01 & 7& 1.31$\pm$  0.20&  11.30$\pm$  0.70& 9.30$\pm$  0.70\\
SAI 104&            & 09:09:54.8& -48:51:03 & 3& 1.17$\pm$  0.11&   9.59$\pm$  0.14& $<$8.90\\
SAI 105&            & 09:10:24.5& -49:54:46 & 4& 1.91$\pm$  0.31&  12.35$\pm$  0.13& 9.05$\pm$  0.05\\
SAI 106&            & 09:17:59.4& -51:01:46 & 5& 1.02$\pm$  0.02&  13.56$\pm$  0.10& 9.35$\pm$  0.05\\
SAI 107& Teutsch 103& 09:28:35.8& -46:35:60 & 3& 0.62$\pm$  0.06&  12.23$\pm$  0.38&$<$ 8.75\\
SAI 108&            & 09:38:52.9& -52:57:28 & 6& 0.66$\pm$  0.02&  12.32$\pm$  0.08& 8.50$\pm$  0.05\\
SAI 109&            & 09:43:34.4& -50:47:33 & 5& 0.23$\pm$  0.04&  14.43$\pm$  0.50& 9.20$\pm$  0.20\\
SAI 110& FSR 1509   & 09:48:00.8& -55:04:01 & 2& 1.17$\pm$  0.13&  11.59$\pm$  0.05& 8.75$\pm$  0.05\\
SAI 111& FSR 1521   & 09:55:22.6& -56:36:15 & 8& 0.90$\pm$  0.17&  13.00$\pm$  0.30& 9.45$\pm$  0.10\\
SAI 112&            & 10:21:24.9& -59:05:34 & 2& 1.00$\pm$  0.02&  14.30$\pm$  0.15& 9.10$\pm$  0.10\\
SAI 114& FSR 1555   & 10:48:55.7& -59:02:58 & 3& 0.76$\pm$  0.05&  13.40$\pm$  0.30& 9.25$\pm$  0.10\\
SAI 115& FSR 1586   & 11:22:44.9& -62:19:06 & 5& 1.60$\pm$  0.29&  11.92$\pm$  0.20& 8.65$\pm$  0.10\\
SAI 116&            & 11:49:18.0& -62:13:47 & 5& 0.98$\pm$  0.08&  11.72$\pm$  0.17& 8.60$\pm$  0.05\\
SAI 117& FSR 1663   & 13:41:04.5& -60:12:27 & 4& 1.01$\pm$  0.16&  11.66$\pm$  0.25& 8.90$\pm$  0.05\\
SAI 118&            & 13:43:03.8& -63:09:53 & 8& 0.17$\pm$  0.21&  10.28$\pm$  0.62& 9.75$\pm$  0.25\\
SAI 120&            & 13:58:42.0& -61:40:08 & 2& 1.99$\pm$  0.30&  12.03$\pm$  0.21& 8.75$\pm$  0.10\\
SAI 121& FSR 1686   & 14:40:06.8& -60:23:13 & 5& 1.48$\pm$  0.23&  13.02$\pm$  0.16& 8.95$\pm$  0.05\\
SAI 122&            & 15:00:03.9& -58:48:13 & 8& 2.26$\pm$  0.43&  11.12$\pm$  0.04& 8.25$\pm$  0.15\\
SAI 123&            & 16:08:17.4& -50:32:06 & 7& 1.35$\pm$  0.18&  11.35$\pm$  0.18& 9.20$\pm$  0.05\\
SAI 124& FSR 1744   & 16:51:35.8& -42:25:47 & 3& 2.58$\pm$  0.34&  12.48$\pm$  0.26& 8.85$\pm$  0.20\\
SAI 127& FSR 124    & 19:06:52.8& +13:14:44 & 4& 0.98$\pm$  0.12&  11.56$\pm$  0.41& 9.10$\pm$  0.05\\
SAI 128& FSR 133    & 19:29:46.9& +15:33:51 & 6& 2.26$\pm$  0.45&  11.59$\pm$  0.12& 8.35$\pm$  0.05\\
SAI 129& FSR 160    & 19:44:44.3& +26:54:08 & 3& 1.42$\pm$  0.05&  14.38$\pm$  0.30& 8.95$\pm$  0.05\\
SAI 130& FSR 158    & 19:47:58.1& +26:01:55 & 8& 1.17$\pm$  0.04&  12.78$\pm$  0.11& $<$ 8.30 \\
SAI 131& FSR 154    & 19:48:00.8& +23:20:53 & 3& 0.70$\pm$  0.13&  12.23$\pm$  0.05& 9.05$\pm$  0.05\\
SAI 132&            & 19:57:01.5& +31:36:31 & 8& 0.94$\pm$  0.09&  12.28$\pm$  0.06& 8.75$\pm$  0.05\\
SAI 133&            & 20:01:58.7& +31:26:18 & 4& 1.80$\pm$  0.05&  13.94$\pm$  0.16& 7.85$\pm$  0.05\\
SAI 135&            & 20:02:55.9& +24:33:17 & 5& 1.34$\pm$  0.29&  12.89$\pm$  0.06& 9.10$\pm$  0.05\\
SAI 136&Kronberger 36&20:04:35.7& +35:13:03 & 6& 0.76$\pm$  0.18&  11.40$\pm$  0.26& $<$8.40\\
SAI 137&            & 20:49:52.6& +41:15:18 & 6& 0.89$\pm$  0.08&  10.12$\pm$  0.24& $<$8.70\\
SAI 138& FSR 275    & 20:56:43.0& +46:53:15 &10& 0.76$\pm$  0.05&  11.98$\pm$  0.11& 9.25$\pm$  0.05\\
SAI 139& Teutsch 22 & 21:01:40.8& +44:06:43 & 2& 0.86$\pm$  0.13&  12.16$\pm$  0.33& 8.55$\pm$  0.05\\
SAI 140& FSR 282    & 21:02:20.1& +48:06:26 & 4& 0.78$\pm$  0.01&  11.79$\pm$  0.12& 9.05$\pm$  0.05\\
SAI 141&            & 21:03:46.9& +46:58:50 & 3& 0.94$\pm$  0.04&  13.47$\pm$  0.16& 8.60$\pm$  0.05\\
SAI 142& Teutsch 156& 21:03:46.7& +47:12:54 & 3& 0.94$\pm$  0.04&  12.97$\pm$  0.22& 8.95$\pm$  0.05\\
SAI 143&            & 21:10:51.4& +50:18:31 & 3& 1.72$\pm$  0.24&  14.07$\pm$  0.17& 9.25$\pm$  0.05\\
SAI 144& Teutsch 74 & 21:45:43.6& +58:05:03 & 4& 1.05$\pm$  0.17&  12.12$\pm$  0.53& 9.05$\pm$  0.05\\
SAI 145& FSR 336    & 21:50:47.5& +55:16:31 &8& 0.55$\pm$  0.09&  11.67$\pm$  0.37& 8.90$\pm$  0.05\\
SAI 146& FSR 342    & 22:07:38.6& +53:06:09 & 8& 0.27$\pm$  0.01&  11.52$\pm$  0.47& 9.05$\pm$  0.05\\
SAI 147& FSR 394    & 22:55:04.8& +58:42:53 & 5& 0.70$\pm$  0.04&  13.73$\pm$  0.21& 9.40$\pm$  0.05\\
SAI 148& FSR 430    & 23:15:29.5& +64:09:58 & 5& 1.08$\pm$  0.20&  11.93$\pm$  0.05& 8.05$\pm$  0.05\\
SAI 149&            & 23:38:01.7& +60:32:57 & 7& 1.29$\pm$  0.30&  11.87$\pm$  0.07& $<$8.20\\
SAI 150& FSR 441    & 23:41:58.8& +58:32:29 & 5& 0.78$\pm$  0.14&  11.84$\pm$  0.04& 8.55$\pm$  0.05\\
SAI 151& FSR 465    & 23:50:54.5& +66:10:49 & 5& 0.78$\pm$  0.04&  13.60$\pm$  0.10& 9.20$\pm$  0.05\\
SAI 153& FSR 460    & 23:59:07.0& +60:40:40 & 7& 0.86$\pm$  0.19&  13.45$\pm$  0.09& 9.20$\pm$  0.05\\
\hline \multicolumn{8}{l}{}\\
\end{tabular}
\end{table*}


\begin{table}[t]

\vspace{0mm} \centering {{\bf Table 2.} Coordinates of new embedded clusters }\label{meansp}
\footnotesize

\vspace{5mm}\begin{tabular}{l|l|c|c|c} \hline\hline
{Cluster} & Other name&  $\alpha_{J2000}$&$\delta_{J2000}$& d\\

& & {$h$  $m$  $s$} & 
{$\circ$  $\prime$  $\prime \prime$} & 
{$\prime$} \\ 

\hline
SAI 10 &           &  00:43:03.2& +41:37:21&  3\\
SAI 15 &           &  01:49:52.6& +53:54:34&  2\\
SAI 18 &           &  02:21:55.9& +61:07:21&  3 \\
SAI 19 &           &  02:26:31.9& +61:59:33&  6 \\
SAI 20 & FSR 584   &  02:27:04.8& +61:37:40&  6 \\
SAI 22 &           &  02:48:58.1& +60:44:34&  5 \\
SAI 23 &           &  02:54:05.0& +60:39:52&  2 \\
SAI 24 &           &  02:59:25.4& +60:33:58&  6 \\
SAI 28 &           &  03:39:16.7& +55:58:24&  3 \\
SAI 32 & FSR 655   &  03:56:11.8& +53:52:09&  1 \\
SAI 38 & FSR 699   &  04:56:05.6& +47:23:25&  2 \\
SAI 39 & FSR 696   &  04:58:31.6& +47:59:03&  1 \\
SAI 52 & FSR 812   &  05:38:09.1& +31:44:16&  3 \\
SAI 53 & FSR 874   &  05:40:58.3& +20:21:18&  3 \\
SAI 54 & FSR 855   &  05:42:22.4& +22:50:04&  3 \\
SAI 55 & FSR 826   &  05:42:51.2& +28:56:43&  3 \\
SAI 97 & FSR 1432  &  08:52:53.3& -44:12:06&  4 \\
SAI 113&           &  10:22:43.6& -59:30:20&  2 \\
SAI 119& FSR 1662  &  13:44:14.2& -62:04:03&  2 \\
SAI 125& FSR 31    &  18:06:27.8& -21:23:27&  5 \\
SAI 126&           &  18:13:26.7& -17:52:49&  3 \\
SAI 134& FSR 198   &  20:02:26.8& +35:40:34&  4 \\ 
SAI 152& FSR 469   &  23:57:04.4& +65:24:52&  1 \\

\hline \multicolumn{4}{l}{}\\
\end{tabular}
\end{table}

\begin{figure}
 \resizebox{\hsize}{!}{\includegraphics{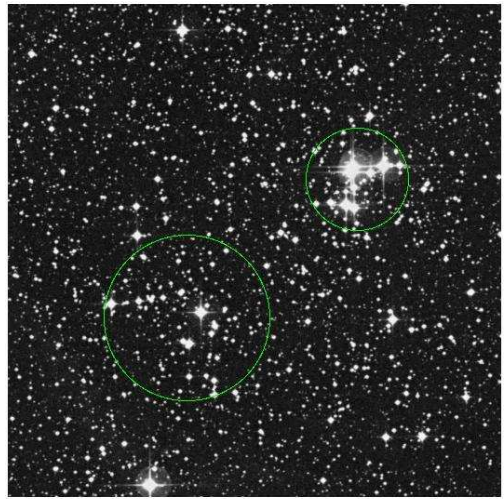}}
\caption{ ~DSS image of double star cluster SAI 92 and NGC 2645 in the field of $15^{\prime} \times 15^{\prime}$. Circles represent the cluster diameters, $5^{\prime}$ and $3^{\prime}$ correspondingly.}
\end{figure}

\section{INVESTIGATION OF YET-UNSTUDIED KNOWN CLUSTERS}

Table 3 lists the coordinates, diameters, color excesses $E(B-V)$, distance moduli, distances in pc, and $log(t)$ of known, but yet-unstudied clusters. We detected and investigated all clusters by our methods, which simultaneously allowed us to improve the accuracy of the center coordinates. For instance, one of the overdensities was attributed to the open cluster Berkeley 53, although it is situated $10^\prime$ away to the west from the coordinates contained in the catalog by Dias et al. (2002). WEBDA gives the same position and the DSS image shows no star concentration there. Hence, Be 53 was not studied previously.

WEBDA misses all $ESO$ clusters, whereas the Dias' catalog lists their coordinates and diameters only. $ESO$ 311-21 is marked by Dias et al. (2002) as ``doubtful'', and $ESO$ 211-03, as ``not found''. Diameter of cluster $ESO$ 368-11 is $1^\prime$ according to Dias et al. (2002), but we found it to be $6^\prime$. Diameters of other known clusters are also less then that in Dias' catalog as compared to the data from Table 3 in this paper. This difference can be explained by the fact that all these clusters are barely perceptible on DSS images, but well noticeable on images from 2MASS. Diameters presented in Table 3 were derived from radial density distribution plots in the same manner as for newly discovered clusters.

Dias' catalog and WEBDA only give the coordinates and underestimated diameters for all $Ruprecht$ clusters from Table 3. Open cluster $BH$ 131 is marked as ``doubtful'' by Dias et al. (2002), $Trumpler$ 19 has no data in the literature available. Koposov et al. (2008) published the coordinates and diameters only for cluster $Koposov$ 7, whereas in present work we determined its age, distance and color excess using the data from UKIDSS GPS.

\begin{table*}[t]

\vspace{0mm} \centering {{\bf Table 3.} Parameters of yet-unstudied clusters}\label{meansp}
\footnotesize

\vspace{5mm}\begin{tabular}{l|l|c|c|c|c|l|c} \hline\hline
{Name} & 
$\alpha_{J2000}$ &
$\delta_{J2000}$& 
{d} & 
{E(B-V)} &  
{$(m-M)_0$} & 
{Distance} & 
{Age}  
\\
&
{$h$  $m$  $s$} & 
{$\circ$  $\prime$  $\prime \prime$} & 
{$\prime$} & 
{$mag$}&
{$mag$}&
pc&
{$log(t)$}
\\
\hline

Koposov 7& 05:40:44.1& +35:55:25 & 6& 3.75$\pm$  0.37&  13.39$\pm$  0.35& 4760$\pm$ 830& 8.20$\pm$  0.05\\
ESO 368-11&07:44:22.2& -34:37:07 & 6& 0.06$\pm$  0.10&  10.86$\pm$  0.41& 1490$\pm$ 310& 9.00$\pm$  0.05 \\
ESO 311-21& 08:01:41.6& -41:35:57 & 5& 0.49$\pm$  0.02&  11.84$\pm$  0.08& 2330$\pm$ 90& 10.10$\pm$  0.05\\
Ru 48 & 08:02:42.5& -32:03:02 & 3& 0.29$\pm$  0.04&  12.41$\pm$  0.05& 3030 $\pm$ 70& $<$8.60\\
Ru 54 & 08:11:21.7& -31:56:37 & 2& 0.35$\pm$  0.04&  12.71$\pm$  0.27& 3480 $\pm$ 460& 8.45$\pm$  0.05\\
Ru 60 &08:24:23.0& -47:12:37 & 4& 0.57$\pm$  0.06&  12.84$\pm$  0.11& 3700 $\pm$ 190& $<$8.65\\
ESO 312-03& 08:24:26.1& -41:17:52 & 6& 0.52$\pm$  0.10&  11.68$\pm$  0.57& 2170 $\pm$ 650& $<$8.90\\
ESO 312-04& 08:26:51.2& -41:14:39 & 2& 0.49$\pm$  0.01&  11.91$\pm$  0.11& 2410 $\pm$ 130& $<$8.75\\
Ru 63 & 08:32:41.4& -48:18:21 & 6& 0.54$\pm$  0.04&  12.13$\pm$  0.08& 2670 $\pm$ 100& $<$8.50\\
Ru 68 & 08:44:34.5& -35:54:00 &10& 0.39$\pm$  0.05&  11.49$\pm$  0.08& 1990 $\pm$ 80& 9.15$\pm$  0.05\\
ESO 211-03& 08:51:34.3& -50:14:43 & 5& 0.90$\pm$  0.07&  12.88$\pm$  0.04& 3770 $\pm$ 70& 9.05$\pm$  0.10\\
TR 19  & 11:14:20.2& -57:33:04 &14& 0.08$\pm$  0.01&  11.40$\pm$  0.05& 1900 $\pm$ 50& 9.65$\pm$  0.05\\
BH 131 & 12:26:13.7& -63:24:50 & 5& 0.62$\pm$  0.04&  13.98$\pm$  0.22& 6250 $\pm$ 670& 9.10$\pm$  0.10\\
Be 53  & 20:55:55.3& +51:02:54 &10& 1.31$\pm$  0.20&  12.74$\pm$  0.13& 3530 $\pm$ 220& 9.15$\pm$  0.05\\

\hline \multicolumn{8}{l}{}\\
\end{tabular}
\end{table*}

\begin{figure}
\resizebox{\hsize}{!}{\includegraphics{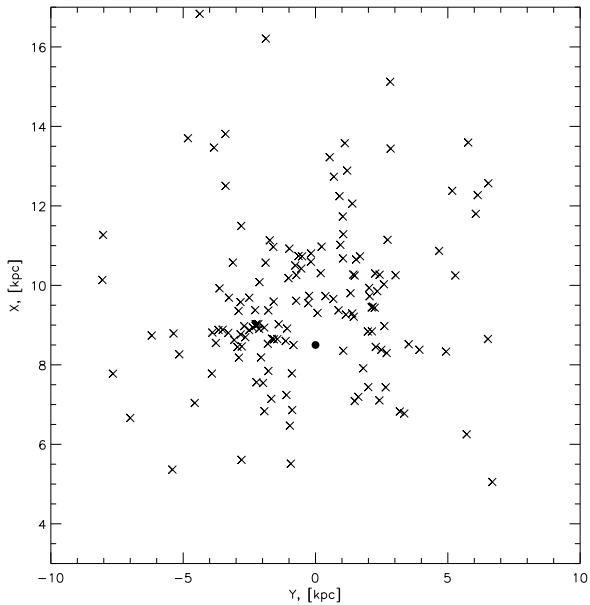}} \caption{Distribution of 143 open clusters across the Galactic plane. Solid circle denotes the Sun.}
\end{figure}

\begin{figure}
\resizebox{\hsize}{!}{\includegraphics{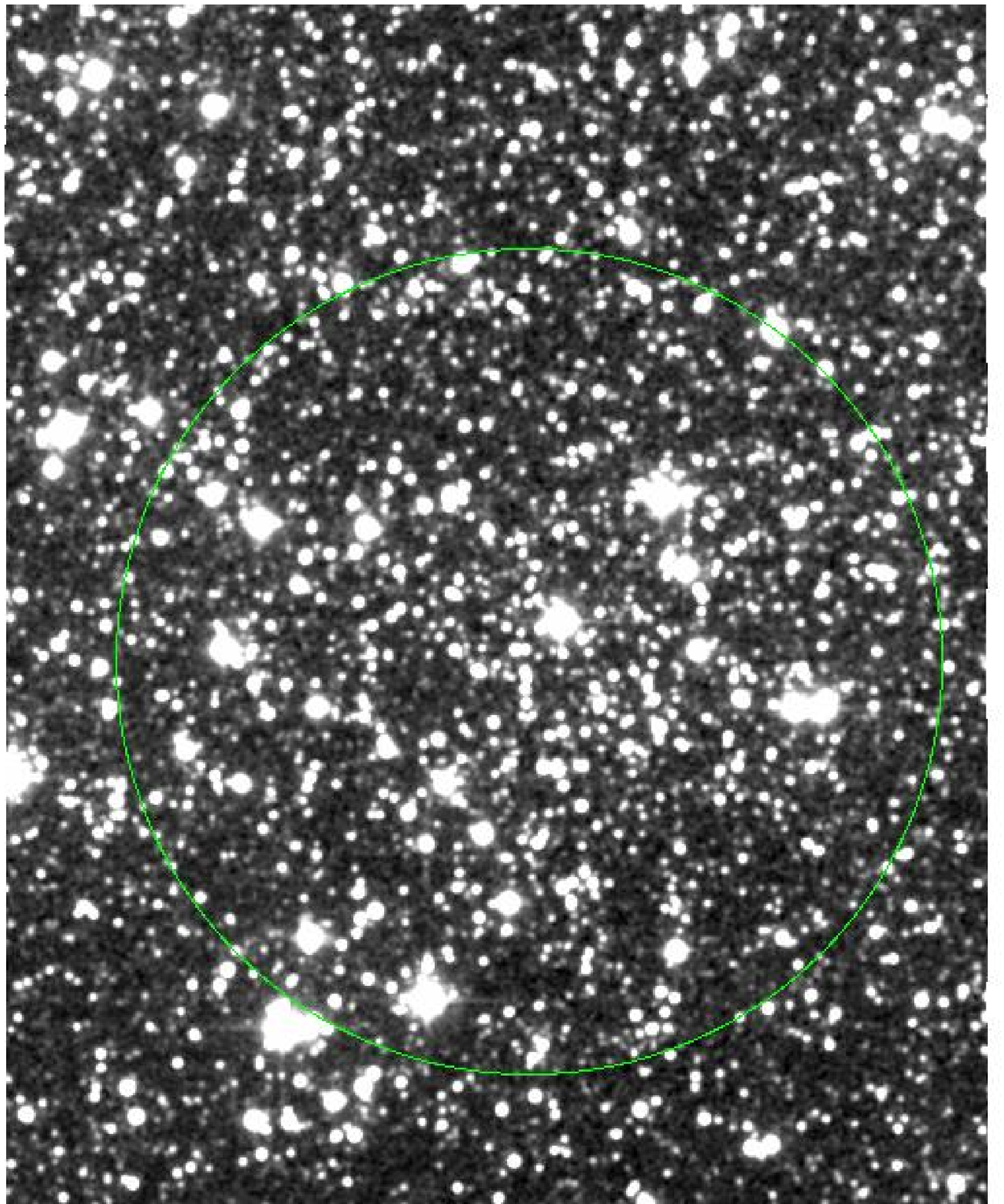}} \caption{2MASS image of cluster SAI 122, diameter is $8^{\prime}$.}
\end{figure}

\begin{figure}
\resizebox{\hsize}{!}{\includegraphics{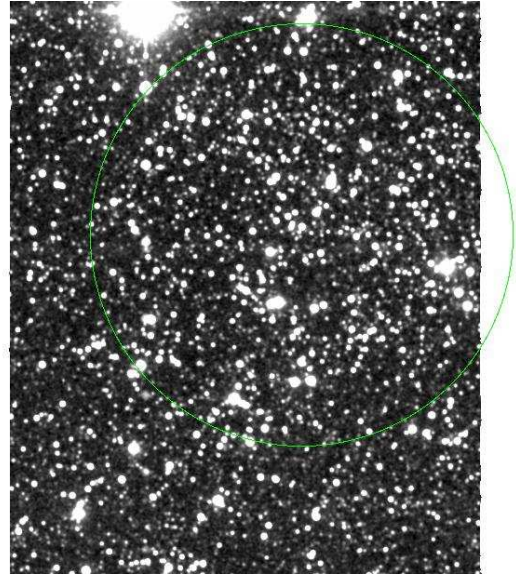}} \caption{2MASS image of cluster SAI 132, diameter is $8^{\prime}$.}
\end{figure}

\section[]{SAI OPEN CLUSTERS CATALOG}

To facilitate dissemination and scientific usage of the results of the present study and data from Koposov et al. (2008), we have developed a standardized tool for continuous publication of the results of ongoing catalog compilation on the Web, available as a dedicated web-site at \url{http://ocl.sai.msu.ru} . Besides the standard access via a web browser to the catalog of individual cluster pages with Hess-diagrams, color-magnitude diagrams, 2MASS or DSS cluster images, and catalogs of photometric data (for clusters with CCD observations), the web-site offers the catalog exported to CSV, DAT, and VOTable formats as well and implements a standard Virtual Observatory programmatic access interface for positional queries called {\it ConeSearch}. Moreover, it is possible to start the VO client software, such as {\sc topcat}\footnote{\url{http://www.star.bris.ac.uk/~mbt/topcat/}} or {\sc cds aladin}\footnote{\url{http://aladin.u-strasbg.fr/}}, by a single click in a web browser with the catalog preloaded for quick-look analysis of the whole sample and all accompanying data we provide for individual clusters.

Main results are available as a single table at Centre de Donnees astronomiques de Strasbourg at \url{http://vizier.u-strasbg.fr/viz-bin/VizieR?-source=V/132} and also within standalone VO client applications running on any workstation connected to the Internet, e.g. mentioned {\sc topcat} and {\sc cds aladin} tools, using query to search for the catalogs by key words or by authors of this publication.

\section{Conclusions}
Using $J,H,K_s$ photometric data available in 2MASS catalog, we have performed an automated search for the star overdensities in the Galactic plane for the latitudes ranged within $|b|<24^{\circ}$ and detected 11186 density peaks. A short list of 283 new cluster candidates has been compiled after we excluded the overdensities attributed to the background fluctuations or star concentrations coinciding with already known clusters. Employing our method described earlier (Koposov et al., 2008), we have studied these 283 candidates and verified 149 of them to be real clusters. Having involved the UKIDSS GPS survey in our investigation as an additional data source, we studied 22 more cluster candidates and found four new open clusters. Therefore, the number of new clusters found in the frame of the present paper is 153. For 130 clusters of them, we have determined their main physical parameters: distance, age, and color excess. Together with the clusters published in Koposov et al. (2008), we have found 168 new open clusters in total, and determined parameters for 142 of them. Besides that, we confirmed the existence of 14 known, but yet-unstudied clusters and determined their accurate coordinates and other parameters.
Fig.~3 shows the distribution of 143 open clusters projected onto the Galactic plane. The crosses denote all clusters, except for SAI 50, whose distances are determined in this paper (129 new and 14 known clusters). We excluded cluster SAI 50 from this pattern because of large errors in its parameters. Solid circle represents the Sun. It can be clearly seen from this figure that no new clusters were discovered in the direction towards the Galactic center, because this area features maximum extinction and level of background fluctuations, which makes the detection of clusters more difficult. The major part of new clusters have been found within the radius of 2.5 kpc from the Sun, because of a relatively low limiting values of $J, H, K_s$ in the 2MASS catalog. We have not discovered any new clusters closer than 500 pc; however, we did not attempt to search for them specifically in this range of the distance, because the open cluster sample there is believed to be complete. For this reason, we used the filter allowing to detect clusters with the diameter of $10^\prime$ and less, whereas for the close clusters this value is typically greater.

Some new rich clusters can only be seen at the IR wavelengths. For example, SAI 122 is well noticeable on the image from 2MASS data (Fig.~4) and is fully invisible on the DSS images, because the color excess in the direction toward this cluster amounts to $2^m.26$. Among the clusters discovered by us, there are few, which reveal themselves on the DSS images as a very faint objects; however, it is normally impossible to estimate their parameters. The majority of new objects are either poorly populated clusters, or rich ones, yet imperceptible against a dense background: for example, SAI 132 (see Fig.~5). In both these cases, an object cannot be detected while visually inspecting 2MASS images. Thus, the application of our automated method of searching infrared data for density peaks has provided us with the tool capable of detecting these 168 open clusters.

\acknowledgements

This work has made use of WEBDA database, operated at the Institute for Astronomy of the University of Vienna and developed by E. Paunzen and J.-C. Mermilliod.

This publication makes use of data products from the Two Micron All Sky Survey, which is a joint project of the University of Massachusetts and the Infrared Processing and Analysis Center/California Institute of Technology, funded by the National Aeronautics and Space Administration and the National Science
Foundation.

This work is based in part on data obtained as part of the UKIRT Infrared Deep Sky Survey.

This publication has made use of the Digitized Sky Surveys which were produced at the Space Telescope Science Institute under U.S. Government grant NAG W-2166. The images of these surveys are based on photographic data obtained using the Oschin Schmidt Telescope on Palomar Mountain and the UK Schmidt Telescope.

This research has made use of the SAI Catalog Access Services, Sternberg Astronomical Institute, Moscow, Russia.

The work was supported by the Russian Foundation for Basic Research (grant no. 08-02-00381).

We are glad to thank Fran\c{c}ois Ochsenbein (CDS) for the comments that helped us to improve the catalog presentation.

\end{document}